\documentclass[twocolumn]{aastex62}

\received{XXX}
\revised{YYY}
\accepted{ZZZ}
\submitjournal{ApJL}

\usepackage{graphicx}	% Including figure files
\usepackage{amssymb}	% Extra maths symbols
\usepackage[inline]{enumitem}
\usepackage{gensymb}

\shorttitle{The Fate of Binaries in the Galactic Center}
\shortauthors{Stephan et al. 2019}
\begin{document}

\title{The Fate of Binaries in the Galactic Center: The Mundane and the Exotic}

\correspondingauthor{Alexander P. Stephan}
\email{alexpstephan@astro.ucla.edu}

\author[0000-0001-8220-0548]{Alexander P. Stephan}
\affil{Department of Physics and Astronomy, University of California, Los Angeles, Los Angeles, CA 90095, USA}
\affiliation{Mani L. Bhaumik Institute for Theoretical Physics, University of California, Los Angeles, Los Angeles, CA 90095, USA}

\author[0000-0002-9802-9279]{Smadar Naoz}
\affiliation{Department of Physics and Astronomy, University of California, Los Angeles, Los Angeles, CA 90095, USA}
\affiliation{Mani L. Bhaumik Institute for Theoretical Physics, University of California, Los Angeles, Los Angeles, CA 90095, USA}

\author[0000-0003-3230-5055]{Andrea M. Ghez}
\affiliation{Department of Physics and Astronomy, University of California, Los Angeles, Los Angeles, CA 90095, USA}

\author[0000-0002-6753-2066]{Mark R. Morris}
\affiliation{Department of Physics and Astronomy, University of California, Los Angeles, Los Angeles, CA 90095, USA}

\author[0000-0001-5800-3093]{Anna Ciurlo}
\affiliation{Department of Physics and Astronomy, University of California, Los Angeles, Los Angeles, CA 90095, USA}

\author[0000-0001-9554-6062]{Tuan Do}
\affiliation{Department of Physics and Astronomy, University of California, Los Angeles, Los Angeles, CA 90095, USA}

\author[0000-0001-5228-6598]{Katelyn Breivik}
\affiliation{Canadian Institute for Theoretical Astrophysics, University of Toronto, 60 St. George Street, ON M5S 3H8, Canada}

\author{Scott Coughlin}
\affiliation{Physics and Astronomy, Cardiff University, Cardiff, CF10 2FH, UK}
\affiliation{Center for Interdisciplinary Exploration \& Research in Astrophysics (CIERA), Northwestern University, Evanston, IL 60208, USA}

\author[0000-0003-4175-8881]{Carl L. Rodriguez}
\affiliation{Pappalardo Fellow; MIT-Kavli Institute for Astrophysics and Space Research, 77 Massachusetts Avenue, 37-664H, Cambridge, MA 02139, USA}

%\author[0000-0003-0395-9869]{B. Scott Gaudi}
%\affiliation{Department of Astronomy, The Ohio State University, Columbus, OH 43210, USA}



\begin{abstract}
    The Galactic Center (GC) is dominated by the gravity of a super-massive black hole (SMBH), Sagittarius A$^*$, and is suspected to contain a sizable population of binary stars. Such binaries form hierarchical triples with the SMBH, undergoing Eccentric Kozai-Lidov (EKL) evolution, which can lead to high eccentricity excitations for the binary companions' mutual orbit. This effect can lead to stellar collisions or Roche-lobe crossings, as well as orbital shrinking due to tidal dissipation. In this work we investigate the dynamical and stellar evolution of such binary systems, especially with regards to the binaries' post-main-sequence evolution. We find that the majority of binaries ($\sim75$\%) is eventually separated into single stars, while the remaining binaries ($\sim25$\%) undergo phases of common-envelope evolution and/or stellar mergers. These objects can produce a number of different exotic outcomes, including rejuvenated stars, G2-like infrared-excess objects, stripped giant stars, Type Ia supernovae (SNe), cataclysmic variables (CVs), symbiotic binaries (SBs), or compact object binaries. We estimate that, within a sphere of $250$ Mpc radius, about $7.5$ to $15$ Type Ia SNe per year should occur in galactic nuclei due to this mechanism, potentially detectable by ZTF and ASAS-SN. Likewise we estimate that, within a sphere of $1$ Gpc$^3$ volume, about $10$ to $20$ compact object binaries form per year that could become gravitational wave sources. {Based on results of EKL-driven compact object binary mergers in galactic nuclei by \citet{Hoang+2018}, this compact object binary formation rate translates to about $15$ to $30$ events per year detectable by Advanced LIGO.}
\end{abstract}


\section{Introduction}\label{sec:intro}

The Galactic Center (GC) contains the closest known super-massive black hole (SMBH), Sagittarius A$^*$, which dominates the gravitational dynamics of its environment due to its large mass of about $4\times10^6~M_{\odot}$ \citep[e.g.,][]{Ghez+2005,Gillessen+2009}. Thus, the environment of the GC has served as a ``laboratory'' to test the nature of gravity, stellar cluster dynamics, and general relativity (GR) over the last decades \citep[e.g.,][]{Ghez+2003,Ghez+2005,Alexander2005,Hopman+2006,Hopman2009,Gillessen+2009,Gillessen+2012,AlexPfuhl2014,Hees+2017,Chu+2018}. The gravitational influence of the SMBH is suspected to cause a number of interesting astrophysical phenomena, such as hypervelocity stars \citep[e.g.,][]{Ginsberg+2007} and stellar binary mergers \citep[e.g.,][]{Antonini2010,Antonini+2011,Prodan+2015,Stephan+2016}. Furthermore, there appears to be a large number of X-ray sources associated with the GC, sometimes interpreted to indicate a large number of X-ray binaries, stellar mass black hole-star binary pairs, where the black hole accretes material from its companion, emitting X-ray radiation \citep[e.g.,][]{Muno+2005,Cheng+2018,Zhu+2018}. \citet{Muno+2006,Muno+2009} and \citet{Heinke+2008}, however, suggested that many of the observed GC X-ray sources are actually Cataclysmic Variables (CVs), consisting of White Dwarf (WD)-main-sequence star binary pairs.

Any binary that orbits the GC's SMBH will feel gravitational perturbations on the binary's orbit by the SMBH. These perturbations can lead to chaotic orbital eccentricity and inclination excitations, the so-called Eccentric Kozai-Lidov (EKL) mechanism \citep[see, for review,][]{Naoz2016}. This effect can cause binary stars in the GC to merge \citep[e.g.,][]{Antonini2010,Antonini+2011,AntPer2012,Witzel+2014,Prodan+2015,Stephan+2016,Witzel+2017}, which has been used to explain extended infrared-excess objects such as G2 \citep[e.g.,][]{Gillessen+2012}. However, the exact nature of these merger-candidate events is not always the same. While some binary stars might just collide or have grazing encounters due to the induced large eccentricities, others might first undergo orbit shrinking and circularization due to tidal dissipation, and might not begin merging until the stars begin to expand due to stellar evolution \citep{Stephan+2016}. The shrinking of the orbit can furthermore ``harden'' the orbit and make it long-term stable against scattering interactions with other stars in the GC \citep[compare also with][]{Trani+2018}, allowing the binary companions to remain coupled long enough for the stars to leave the main-sequence and become red giants. 

In this work, we investigate the possible outcomes of post-main-sequence binary star evolution in the GC, taking into consideration the effects of EKL, tides, GR, and interactions with the GC's stellar cluster. Most of these stars end their regular lives as WDs ($M_{*,ini}\lesssim 8~M_{\odot}$), as these constitute the bulk of the stellar population \citep[e.g.,][]{Salpeter1955}, and which could serve as progenitors of Type Ia supernovae (SNe). Furthermore, we also consider progenitors of stellar-mass black holes and neutron stars. 

\section{Numerical Setup}\label{sec:num}

\begin{figure}
\hspace{0.0\linewidth}
\includegraphics[width=\linewidth]{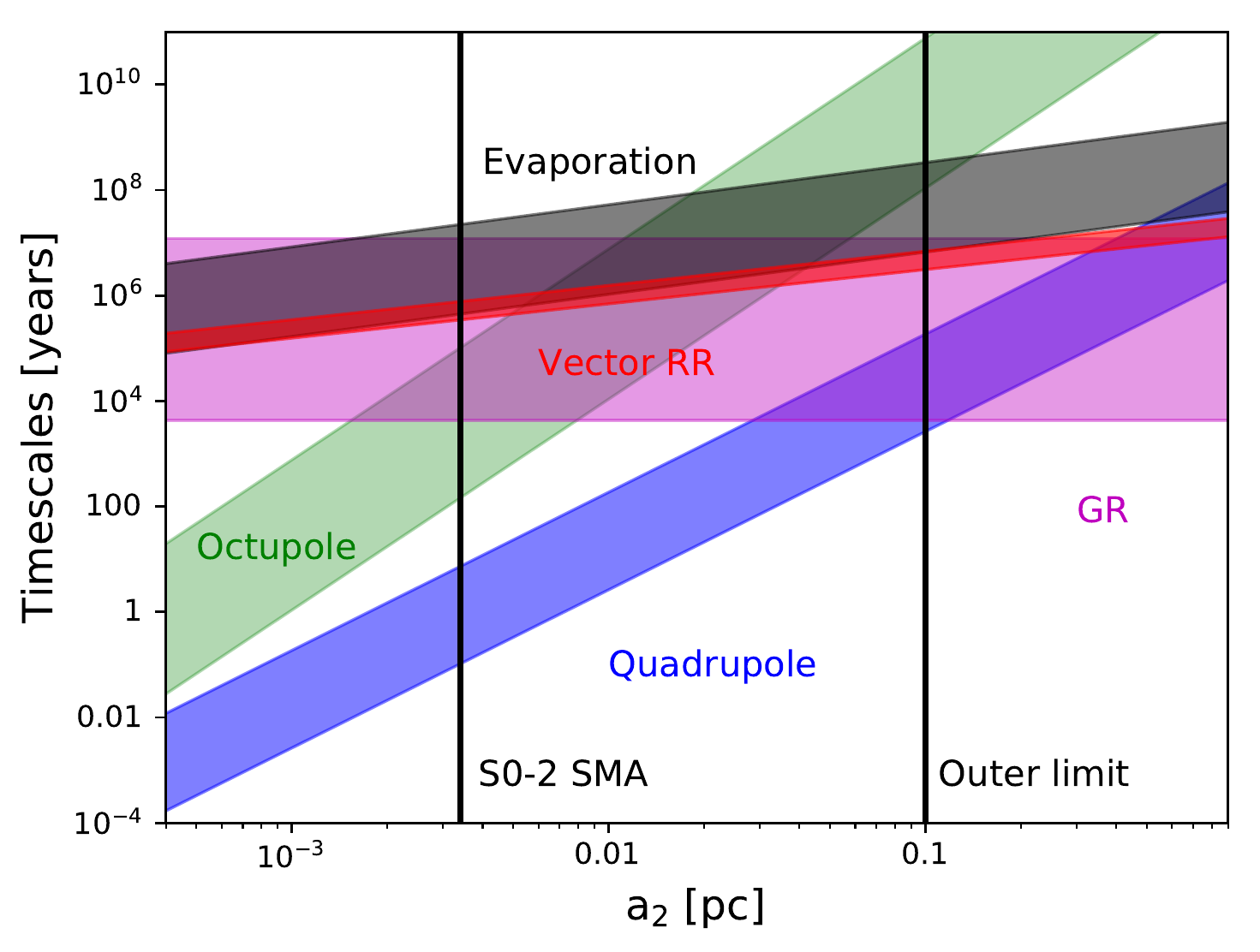}
\caption{{\bf Timescales of physical effects in the GC.} The figure shows the timescales for several physical effects acting on binaries in the GC as a function of outer orbital distance from the SMBH, such as GR (magenta), vector resonant relaxation (red), evaporation (dark grey), and the timescales for the quadrupole (blue) and octupole (green) levels of the EKL mechanism. The filled areas show the strengths of these effects for a range of possible binary parameters, ranging in combined binary mass from $2$~to $10$~M$_\odot$ and in binary semi-major axis from $1$~to $10$~AU. The black lines show the inner and outer limits of semi-major axis (SMA) values around the SMBH used in this work, from $\sim700$~AU to $0.1$~pc.}
   \label{fig:masslifetime}
\end{figure}

In \citet{Stephan+2016}, we executed a large Monte-Carlo simulation of binary stars in the inner $0.1$~pc around the GC's SMBH that explored the dynamical evolution of the binaries until they either merged, tidally locked, or separated. In this work, we expand on these earlier simulations and use the same system parameters for our new Monte-Carlo runs: the primary stellar mass was chosen from a Salpeter distribution with $\alpha=2.35$ \citep{Salpeter1955}, with the masses limited between $1$ and $150~M_{\odot}$, the mass ratio to the secondary was taken from \citet{Duquennoy+91}, and the mass of the SMBH was set to $4\times10^6~M_{\odot}$ \citep[e.g.,][]{Ghez+2005,Gillessen+2009}. The inner binary semi-major axis distribution was also taken from \citet{Duquennoy+91}, while the outer binary semi-major axis was initially drawn uniformly between about $700$~AU and $0.1$~pc. The inner limit corresponds approximately to the orbit of the star S0-2 \citep{Ghez+2005}, while for most orbits beyond $0.1$~pc the effects of vector resonant relaxation become non-negligible \citep[e.g.,][]{Hamers+2018}. This is also shown in Figure \ref{fig:masslifetime}, where we show several important dynamical timescales for the GC. The inner binary orbit eccentricity was drawn uniformly between $0$ and $1$ \citep{Raghavan+10}, while the outer one was taken as thermal \citep{Jeans1919}. The angle between the inner and outer angular momenta was drawn isotropically. These systems were then tested for orbital stability, requiring that 
\begin{equation}\label{eq:eps}
    \epsilon =\frac{a_1}{a_2}\frac{e_2}{1-e_2^2},
\end{equation}
where $a_1$ ($a_2$) is the inner (outer) orbit semi-major axis, and $e_2$ is the outer orbit eccentricity. Further analytic stability criteria had to be fulfilled, ensuring that the binaries do not cross the SMBH's tidal radius \citep[see][for a complete list of conditions]{Stephan+2016,Naoz2016}.
In total, we had $5{,}203$ stable systems, $1{,}570$ from \citet{Stephan+2016} and $3{,}633$ additional ones, to provide better statistical significance. The distribution of semi-major axis and eccentricity values after application of the stability criteria for the new simulations are equivalent to the old ones shown by \citet{Stephan+2016}, Fig. 2.

The systems were then evolved including effects from EKL, GR, tides \citep[e.g.,][for the complete set of equations]{Naoz2016}, and stellar evolution, following the single-star stellar evolution code {\tt SSE} by \citet{Hurley+00}\footnote{It has been shown that stellar evolution can have significant effects on the dynamical evolution in many astrophysical settings involving the EKL mechanism \citep[e.g.,][]{Shappee+13,Petrovich+2017,Stephan+2016,Stephan+2017,Stephan+2018}.}. The time it takes on average for binaries to be separated by interactions with other stars in the GC, (called the ``evaporation timescale'') served as the time limit for our calculations. However, if tidal effects are able to shrink the inner orbital distance of a binary, its expected survival time increases, as it becomes more stable against such scattering events \citep[for equations, see][]{Binney+Tremaine1987, Stephan+2016}. {Scattering events can also lead to exchanges of stars in the binaries with other objects orbiting the GC \citep[e.g.,][]{Trani+2018}; while this can have interesting implications for further secular binary evolution, the encounter and exchange rate for tidally shrunken binaries is rather low and does not substantially influence our results.} Furthermore, short-period binaries do not feel strong EKL effects, as GR and tidal effects suppress the gravitational perturbations by the SMBH. This leads to the formation of a large population of long-lived short-period binaries, as described by \citet{Stephan+2016} and in the present work. 

High-eccentricity mergers induced by EKL occur very early in the binaries' lifetime \citep[as shown by][]{Stephan+2016}. Radial mergers are caused by the radial expansion of one star in the binary and do not necessarily require EKL-induced eccentricity or tidal orbital shrinking (though either effect facilitates the occurence of such mergers), and therefore occur somewhat later, when the most massive stars begin to evolve past the main-sequence. Tidal mergers are binaries that become short-period binaries due to the interplay of EKL-induced high eccentricities and tidal orbital shrinking, destined to merge during post-main-sequence stellar evolution due to stellar expansion, after hundreds of Myrs or even several Gyrs. Tidal mergers were the dominant merging mechanism described by \citet{Stephan+2016}, producing $56$\%~of mergers, compared to $33$\%~of mergers through EKL during the main-sequence, and $11$\%~of mergers due to radial expansion of evolved, massive stars.

We use the merging binary systems from our simulations as the basis for this work. We take the orbital parameters of these binaries at the moment when they either crossed the Roche lobe or became tidally locked short-period binaries and determine their further binary star evolution using the binary stellar evolution code {\tt BSE} \citep{Hurley+02}. In particular, we use the {\tt BSE} version distributed with the {\tt COSMIC} binary population synthesis suite.

Several updates have been made to the {\tt BSE} version used in {\tt COSMIC}. Metallicity-dependent wind prescriptions for high-mass stars have been implemented following \cite{Vink2001},\cite{Vink2005}, and \cite{Belczynski2010}. Prescriptions for compact object formation and natal kicks have also been updated following \cite{Fryer2001}, \cite{Fryer2012}, and, for electron-capture SNe for neutron stars, following \cite{Kiel+2008}. We used the default parameters for {\tt BSE} except in the case of the updated prescriptions detailed above. In particular, we assume the `rapid' model from \cite{Fryer2012} for compact object formation and that natal kicks are drawn from a Maxwellian distribution with $\sigma=265\,$km/s \citep{Hobbs2005}, with BH natal kicks being modified due to the amount of mass that falls back onto the proto-BH during formation. Finally, we note that our treatment for the common envelope follows the standard $\alpha\lambda$ formalism and default {\tt BSE} values, with $\alpha=1.0$ and $\lambda$ determined by the stellar properties at the time just prior to the common envelope \citep{Claeys2014}.

{\tt BSE} keeps track of the stars' evolutionary phases and, furthermore, indicates when and if binary stars merge and what type of object would be the outcome of such mergers. Double WD mergers would be candidates for Type Ia SNe, if their combined mass is large enough, while WD-red giant and WD-main-sequence mergers are candidates for symbiotic binaries (SBs) and CVs, as the WD would easily be able to accrete material from the companion star's envelope. Mergers involving helium stars, red giant stars, and main-sequence stars are probably candidates for G2-like objects, stars shrouded by a gaseous dust-rich cloud for a few Myrs after merging had occurred \citep{Witzel+2014,Witzel+2017,Stephan+2016}. Helium stars themselves are examples of ``stripped giants'', which have been speculated to exist in the GC \citep{Ghez+2008}.

We note that while we use {\tt BSE} to perform the calculations for merger candidates' further stellar evolution, even for eccentric binaries, there are some limitations with this treatment \citep{Sepinsky+2007,Sepinsky+2007a,Sepinsky+2010}. However, {\tt BSE} does include effects like SN kicks for neutron stars and provides results massively faster than full stellar hydrodynamic models. These are important advantages that allow us to advance this work.

\section{Results and Discussion}\label{sec:res}

Here we go into greater detail about the different types of outcomes and their implications for the binary population in the GC.

\subsection{Dynamical Evolution Outcomes}\label{subsec:dyn}

We summarize the dynamical evolution outcomes before including binary evolution calculations below.

\begin{enumerate}

    \item {\bf Unbound Binaries:} $75$\% of the initial binary population is separated by interactions with other stars in the GC into single stars before the binary members can interact in a meaningful way. This unbinding (evaporation) of binaries begins after a few Myrs of dynamical evolution and is mostly finished after a few $100$~Myrs ($\sim25$\% of binaries have separated after about $6$~Myrs). After that point, any binary that has not separated has either merged or has a tidally shrunken orbit that is long-term stable against separation.
    
    \item {\bf EKL Mergers:} $10$\% of binaries merge due to high eccentricities induced by EKL-oscillations. The binaries merge early, within the first few Myrs of their dynamical evolution.
    
    \item {\bf Radial Mergers:} $2$\% of binaries merge simply due to radial expansion of one of their members during the red giant phase. These systems generally include very massive stars that evolve rapidly, within the first few Myrs of evolution.
    
    \item {\bf Tidally Locked:} $13$\% of binaries have tightened and circularized their orbits due to the interplay of EKL and tides to such a degree that they are long-term stable against separation. These systems must reach this tidally locked state within only a few Myrs after formation to escape evaporation. Their subsequent fate then depends on the stellar masses and orbital separation.
    
\end{enumerate}

\begin{figure}
\hspace{0.0\linewidth}
\includegraphics[width=\linewidth]{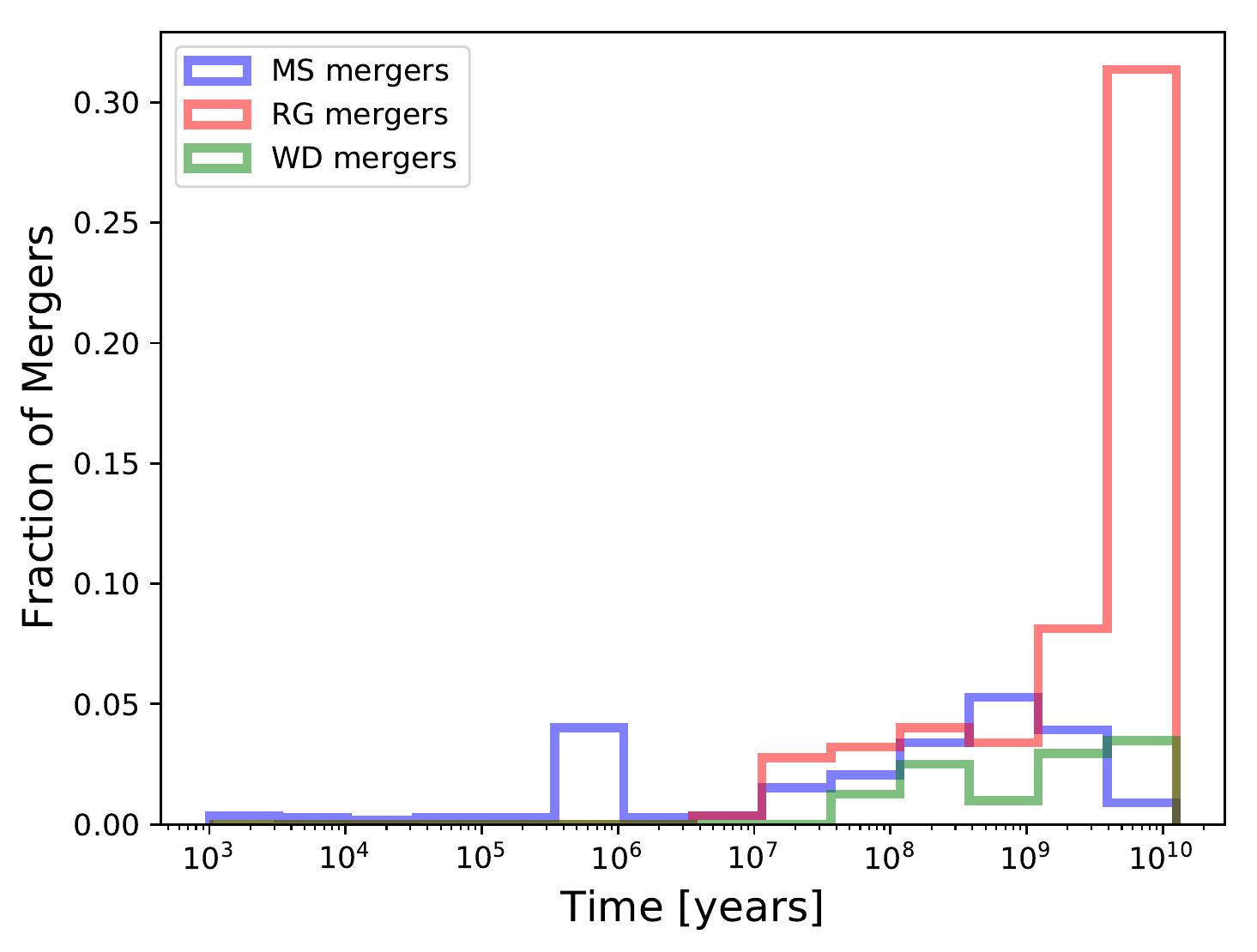}
\caption{{\bf Types of mergers as a function of time.} The plot shows a histogram of the times the main types of mergers occur. Mergers of two main-sequence stars are shown in blue, mergers involving at least one red giant star (and no WD or other compact object) are shown in red, and mergers involving at least one WD are shown in green. The peak of main-sequence binary mergers in the first Myr of evolution is due to EKL-driven high-eccentricity collisions, while the continued merging of RG and WD systems over several Gyrs is due to tidally shrunken and circularized binaries. Mergers that involve at least one BH or NS constitute about $1\%$ of the systems and are not depicted here to avoid clutter.}
\label{fig:histogram}
\end{figure}

\subsection{Binary Evolution Outcomes}\label{subsec:bin}

The different outcomes of binary evolution are essential to understand the census of binaries in galactic nuclei. The outcomes were determined using {\tt BSE}, as applied to EKL and radial mergers and as well as the tidally locked binaries described above. We refer to these systems as ``merger candidates.'' {\tt BSE} evolves these systems while keeping track of common envelope phases, Roche lobe overflow phases, mergers, and kicks and possible separation. Sometimes, this will result in the formation of very tight compact object binaries instead of true mergers \citep[e.g.,][]{TaamRicker2010}, though, it can usually be expected that such systems merge eventually either through gravitational wave emission or further gravitational perturbations by the SMBH. In general, the results can be divided into three groups, based on the evolutionary state of the stars when they merge:

\begin{figure*}
\hspace{0.0\linewidth}
\includegraphics[width=\linewidth]{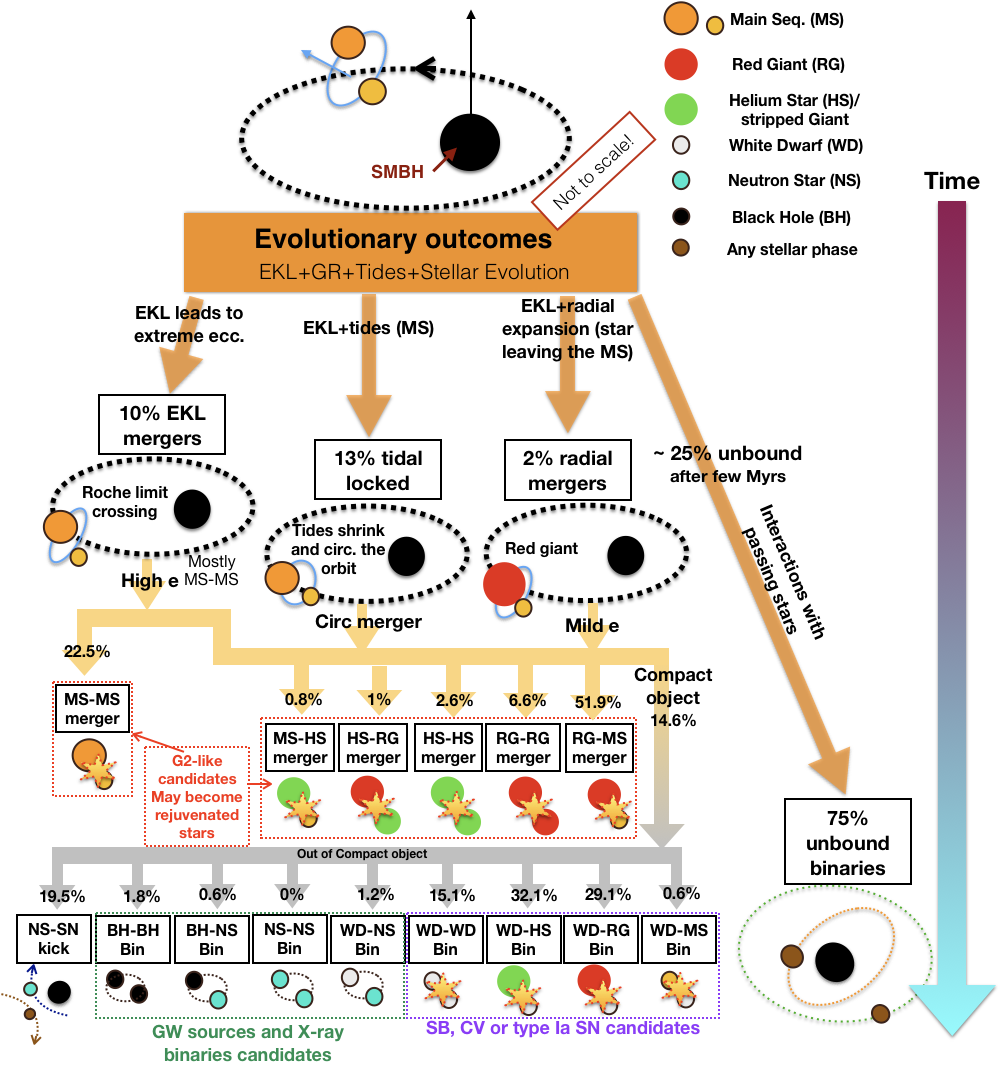}
\caption{{\bf Flowchart of binary evolution outcomes.} The diagram shows the outcomes of binary evolution in the inner $0.1$~pc of the GC. Dynamical effects such as scattering with other stars separate $75$\% of all binaries into independent singles before they can interact with each other. $10$\% of binaries will collide or have grazing encounters due to EKL-induced high eccentricities. $2$\% will merge simply due to radial expansion of one of the binary members due to stellar evolution. $13$\% of binaries will tidally shrink their orbits and become decoupled from gravitational perturbations by the SMBH. We determined the further evolution of these binary pairs and their evolutionary phases during merging using {\tt BSE}. The different possible outcomes are shown in the figure. Generally, the most likely combinations for merging binaries are pairs of main-sequence stars, main-sequence and red giant stars, pairs of red giant stars, and white dwarfs with evolved stellar companions. There is also a sizable population of binaries that were separated due to neutron star kicks, producing single neutron stars orbiting the SMBH, as well as a small population of binaries containing black holes or neutron stars that can become gravitational wave sources and might have been X-ray sources at some point, marked by the green box \citep[see][]{Bortolas+2017,Hoang+2018}. Mergers involving white dwarfs are candidates for SBs, CVs, and Type Ia SNe (purple box), while mergers involving red giants, stripped giants, or main-sequence stars are candidates for G2-like objects or progenitors of rejuvenated stars (red boxes) \citep[e.g.,][]{Witzel+2014,Witzel+2017,Stephan+2016}.}
\label{fig:flowchart}
\end{figure*}

\begin{enumerate}

\item {\bf Main-sequence or Red Giant Mergers:} $85.4$\% of merger candidates eventually merge as some combination of main-sequence, red giant, and/or helium (stripped giant) stars. We extrapolate from current observations of binary star mergers \citep[e.g.,][]{Tylenda+2011_1,Tylenda+2011_2,Tylenda+2013,Nicholls+2013,Kaminski+2018} that the forming objects will be shrouded by extended gas and dust clouds for an extended period of time\footnote{Merged stars undergo a much more violent evolution, where it can be expected that a lot of material is ejected and forms an extended envelope. Thus, while the Kelvin-Helmholtz timescale of a new star is on the order of a few $10^5$ years, an extended envelope may still be engulfing it.}. These mergers are therefore candidates for explaining the sizable population of observed G2-like objects \citep{Ciurlo+2019}. We show the occurrence rate of these mergers, split into mergers only involving main-sequence stars and mergers involving red giants, as the blue and red, respectively, histograms in Figure \ref{fig:histogram}. Note that main-sequence star-only mergers tend to occur early, especially due to EKL-driven high eccentricity mergers.
    
The resulting merger product may appear as a rejuvenated star, as it is now a more massive star that has burned less fuel than equally massive stars that evolved as singles. Such stars are also known as blue stragglers in open clusters \citep[e.g.,][]{PF09,aaron+11,Naoz+14stars}. The difference in measured age versus the actual age of the star could be substantial, but is highly dependent on the types of stars and their evolutionary state at the moment of merging, as well as the details of the merging process. According to the results given by {\tt BSE}, the new star could appear younger by several Gyrs, or just a few Myrs, depending on the exact circumstances. Exploring the exact stellar evolution processes that determine the degree of rejuvenation, however, goes beyond the scope of this work. {Regarding mergers involving red giants and helium stars, previous studies suggest that the resulting products can be carbon stars or even R Coronae Borealis (R CrB) variables \citep[e.g.,][]{Iben+1996,Izzard+2007,Zhang+2013}.}

We note here that a small fraction of binaries can produce Type Ia-like SNe during their red giant evolution. These are mostly similar-mass stellar pairs on very short orbits ($\sim1$~day). When these binaries evolve into red giants, they form Helium cores that can eventually collide as the stars enter common-envelope evolution and coalesce into a single object. As they collide, some of these Helium cores are able to ignite and explode the stars without leaving a remnant behind. While this is very similar to a Type Ia SN, the cores' combined mass is generally very low ($\lesssim0.5$~M$_\odot$), {producing probably somewhat less energy than a typical Type Ia SN ($\sim10^{50}-10^{51}$~ergs) \citep[for a review of types of SNe and their luminosities, see][]{Kasliwal2012}; such events might therefore be more difficult to identify and characterize correctly when occurring in the crowded environment of galactic nuclei.} However, the combination of tidal effects and EKL with a SMBH serves to strongly enhance the formation of such short-period binaries. According to {\tt BSE}, about $1$\% of the entire GC stellar binary population lead to such low-energy SNe.
    
\item {\bf White Dwarf Binaries:} $11.2$\% of merger candidate systems evolve to contain a WD star with either a WD, stripped giant, red giant, or main-sequence star companion. Such systems (i.e., WD-WD, or WD-MS) are candidates for Type Ia SNe, given enough mass or enough collision energy. {\tt BSE} evolution concluded that $15.4$\% of these systems would result in Type Ia SNe (which is $0.4$\% of the entire GC stellar binary population). Binaries and higher-multiplicity systems containing WDs outside of the GC have been investigated in previous studies as potential sources for WD collisions and Type Ia SNe, including due to EKL effects \citep[e.g.,][]{Katz+12,Toonen+2017}. The occurrence rate  over time of these mergers is shown in green in Figure \ref{fig:histogram}.

%\footnote{{\tt BSE} predicts Type Ia SNe to occur for a number of WD-WD pairs that have combined masses significantly lower than the Chandrasekhar limit. Furthermore, {\tt BSE} also predicts similar SN events to occur for some RG-RG mergers, which we have not included in our estimated number for Type Ia SNe.}. 

Given our results for Type Ia SNe, we can estimate the occurrence rate of such SNe in the galactic neighborhood. The GC contains on the order of $10^7$~M$_\odot$ of stars and stellar remnants within a radius of about one parsec from the SMBH \citep[e.g.,][]{Genzel+2003,Schoedel+2003}. Given an age of the Galaxy of about $10$~Gyrs, the star formation rate is approximately $10^{3}$~M$_\odot/$Myr. Assuming a Salpeter Initial Mass Function (IMF), and an average star mass of $1$~M$_\odot$, the Type Ia SN likelihood is $0.4$\% and the Type Ia SN rate is $4\times10^{-6}/$yr for our GC \footnote{This is a conservative estimate, as the GC's actual IMF is top-heavy \citep{Lu+2013}. We would expect this to lead to a higher rate than the one calculated here.}. Based on the observation statistics of the ASAS-SN collaboration \citep[e.g.,][]{ASASSN_Holoien+2018}, we assume an effective observation radius of $250$~Mpc (up to redshift $0.06$), or an effectively observable sphere of $0.065$~Gpc$^3$. Assuming a galaxy density of $0.02$~Galaxies$/$Mpc$^3$ \citep{Conselice+2005}, there are about $1.3\times10^6$ galaxies within this sphere. If we assume that half (or all) of these galaxies have central massive BHs and nuclear star clusters similar to ours, the expected rate of Type Ia SNe within this observable sphere would be about $2.5/$yr (or $5/$yr), on average. If we also include Type Ia-like SNe from colliding Helium cores (see red giant mergers), the rate is approximately tripled, to a maximum of $7.5/$yr (or $15/$yr), on average; unfortunately, due to their low energy, these explosions might be more challenging to observe. However, the Zwicky Transient Facility (ZTF) should be capable of observing even these fainter supernovae, at even greater distances \citep{Bellm+2019}. 
    
Furthermore, binary pairs of WDs and main-sequence or red giant stars are also candidates for CVs or SBs, respectively, as the WDs will be accreting material from their companions for some part of their evolution. These objects, as well as some WD pairs, could remain as tight binaries for extended periods of time before merging \citep[e.g.,][]{TaamRicker2010}. The large population of observed X-ray sources in the GC might be explained by these WD harboring binaries \citep[see also][]{Zhu+2018}. {As the binary pairs eventually merge, some of them are candidates for carbon stars or, in some cases, R CrB variables \citep[e.g.,][]{Iben+1996,Izzard+2007,Zhang+2013}.}
    
\item {\bf Black holes and Neutron stars:} $3.4$\% of merger candidates will eventually evolve to contain a stellar mass black hole or a neutron star, {many of which could have been high-mass X-ray binaries (HMXB) before the secondary became a compact object, however most of these pairs will separate due to SN kicks when the neutron stars form \citep[see also][]{Bortolas+2017,Lu+2019}.  \citet{Lu+2019} showed that this can also bring NSs onto eccentric orbits close enough to the SMBH to merge via gravitational wave emissions, which should be detectable by LISA.} Nevertheless, a small number of binary black holes and black hole-neutron star binaries will survive ($\sim0.1$\% of the entire GC binary population) and can serve as gravitational wave sources in the GC \citep[e.g.,][]{Hoang+2018}. The formation rate of such binaries would be about $10^{-6}/$yr for our GC (see calculations above for Type Ia SNe, assuming again a Salpeter IMF and an average star mass of $1$~M$_\odot$). Within a sphere of $1$~Gpc$^3$ volume, with $0.02$~Galaxies$/$Mpc$^3$, and half (up to all) of galaxies containing an MBH and nuclear star cluster, the total expected formation rate of compact object binaries is about $10/$yr to $20/$yr. {Assuming a merger efficiency of these compact object binaries due to further EKL effects of about $10$\%, as was shown by \citet{Hoang+2018}, this formation rate translates to about $1$ to $2$ gravitational wave signal producing inspirals per year per Gpc$^3$. Given Advanced LIGO's detection range of up to $1.5$ Gpc \citep{Abbott+2018}, this implies a rate of about $15$ to $30$ detectable events per year, with $7$ to $15$ ($2$ to $5$) being BH-BH (BH-NS) mergers. Other merger scenarios for BH binaries in galactic nuclei exist \citep[e.g.,][]{Stone+2017,Bartos+2017}, however future observations by LISA should be able to distinguish between them \citep{Hoang+2019}.}
    
{There will also be a small number of neutron stars with WD companions, which can evolve into ultracompact X-ray binaries \citep[e.g.,][]{Bobrick+2017}. Some of these systems will also have been low-mass X-ray binary (LMXB) or transient X-ray source \citep{Zhu+2018} candidates before the secondaries became WDs, potentially explaining part of the observed GC population of X-ray sources \citep[e.g.,][]{Hailey+2018} \citep[for examples of LMXB formation models, see][]{Naoz+2016,Generozov+2018}.}
    
\end{enumerate}

The dynamical and evolutionary outcomes of our simulations listed above are summarized in Figure \ref{fig:flowchart}.

\section{Conclusions}\label{sec:concl}

We have followed the dynamical evolution of binary stars in the vicinity of a SMBH following a burst of star formation, including the effects of stellar evolution, tides, and GR. In the first few Myrs, about $25$\% of binaries became unbound due to scattering interactions with passing stars, while $10$\% merged due to EKL-induced high eccentricities, $2$\% crossed their companion's Roche lobe due to stellar expansion, and $13$\% became tightly bound binaries due to tidal dissipation. Over the next few Gyrs, the fraction of unbound binaries rose to $75$\%, becoming part of the rather mundane population of GC single stars, while the merged and tightly bound binaries evolved into a number of more exotic outcomes. These outcomes include G2-like objects shrouded by gas and dust for an extended period of time, white dwarfs with main-sequence and red giant star companions forming CVs and SBs, some Type Ia SNe from WD-WD pairs, as well as Type Ia-like events from the collision of red giant Helium cores. Furthermore, some compact object binaries form that could be detected as gravitational wave sources. The results are summarized in Figure \ref{fig:flowchart}.

These results, of course, can be extended to more continuous or episodic star formation histories, leading to the continuous formation of such binary merger products. If this is the case, we estimate that Type Ia SN detecting surveys such as ASAS-SN and ZTF would see about $7.5$ to $15$ Type Ia SN, on average, per year from the processes described in this work, out to redshift $0.06$. Furthermore, this study suggests that compact object binaries can form at a rate of about $10$ to $20$~yr$^{-1}$~Gpc$^{-3}$, {which should yield about $15$ to $30$ gravitational wave inspiral events per year detectable by Advanced LIGO, with $7$ to $15$ ($2$ to $5$) being BH-BH (BH-NS) mergers.}

\section*{Acknowledgements}

S.N.~thanks the KECK Foundation for its partial support of the NStarOrbits Project. S.N.~acknowledges the partial support of NASA grant No.~80NSSC19K0321 and also thanks Howard and Astrid Preston for their generous support. S.N.~and A.P.S.~also acknowledge partial support from the NSF through grant No.~AST-1739160. Calculations for this project were performed on the UCLA cluster {\it Hoffman2}. We thank Vicky Kalogera for facilitating the involvement of {\tt COSMIC} team members in this work and for useful discussions in updating the {\tt BSE} version used in {\tt COSMIC}. We thank the anonymous reviewer for their suggestions to improve this paper.

%%%%%%%%%%%%%%%%%%%%%%%%%%%%%%%%%%%%%%%%%%%%%%%%%%



\bibliographystyle{aasjournal}
\bibliography{Kozai2} % if your bibtex file is called example.bib

\end{document}